\documentclass[aps,prd,preprint,superscriptaddress,tightenlines,floatfix,showpacs,11pt]{revtex4-1}
\usepackage{graphics}
\usepackage{graphicx}
\usepackage{amsmath,amssymb}
\usepackage{tabularx}
\usepackage{array}
\usepackage[english,activeacute] {babel}

\begin{document}

\title{Quantum gravity modifications to the accretion onto a Kerr black hole}

\author{Luis A. S\'anchez}\email{lasanche@unal.edu.co}

\affiliation{Departamento de F\'\i sica, Universidad Nacional de Colombia,
A.A. 3840, Medell\'\i n, Colombia}

\begin{abstract}
In the framework of the Asymptotic Safety scenario for quantum gravity, we analyze quantum gravity modifications to the thermal characteristics of a thin accretion disk spiraling around a renormalization group improved (RGI-) Kerr black hole in the low energy regime. We focused on the quantum effects on the location of the innermost stable circular orbit (ISCO), the energy flux from the disk, the disk temperature, the observed redshifted luminosity, and the accretion efficiency. The deviations from the classical general relativity due to quantum effects are described for a free parameter that arises in the improved Kerr metric as a consequence of the fact that the Newton constant turns into a running coupling $G(r)$ depending on the energy scale. We find that, both for rapid and slow rotating black holes with accretion disks in prograde and retrograde circulation, increases in the value of this parameter are accompanied by a decreasing of the ISCO, by a lifting of the peaks of the radiation properties of the disk and by an increase of the accretion mass efficiency, as compared with the predictions of general relativity. Our results confirm previously established findings in Ref.~\cite{r17} where we showed that these quantum gravity effects also occur for an accretion disk around a RGI-Schwarzschild black hole.\\

\noindent
{\bf keywords}: Quantum aspects of black holes, asymptotic safety, accretion disks.
\end{abstract}

\pacs{04.60.-m, 04.70.-s}

\maketitle

\section{\label{sec:intr}Introduction}
Currently it is widely accepted that classical unphysical singularities associated to, e.g., the origin of the universe and those arising inside the event horizon of black holes (BH), can be solved within the framework of a quantum theory of gravity. Among the several theoretical proposals for such a theory, Asymptotic Safety (AS) is distinguished for being based entirely on the standard quantum field theory as it makes use of the powerful techniques of the functional renormalization group (FRG) \cite{r1}. The FRG is a functional non-perturbative evolution equation, the Wetterich equation \cite{r2}, that controls the flow of the scale dependent quantum effective action for gravity. In this context, the AS scenario postulates the existence of an ultraviolet (UV) non-Gaussian fixed point (NGFP) of the flow that realizes quantum scale invariance beyond a microscopic transition scale which is assumed to be the Planck scale. Beyond this scale a finite number of essential physical degrees of freedom become scale-invariant which makes them safe from unphysical divergences at all scales and, consequently, ensures that the theory is UV-complete and predictive up to the highest energies \cite{r3,r4,r5}.\\

\noindent
The ability of AS for solving the problem of the BH singularities present in the classical theory of general relativity (GR) 
lies in the fact that quantum scale invariance beyond the Planck scale leads to the antiscreening character of gravity in the UV, which means a weakening of gravity at high energies. Indeed, it has been shown that quantum effects associated to the collapse of dust in the AS scenario lead to non-singular black hole solutions \cite{r6,r7,r8,r9,r10}. On the other hand, the possibility of observational test of gravity theories in the strong field regime coming from the detection of gravitational waves from binary black hole mergers, has stimulated the study of quantum gravity modifications to classical BH spacetimes, the so-called renormalization group improved (RGI-) metrics, which lead to infrared (IR) quantum effects that manifest themselves at radial distances greater than the BH horizon. Studies in this direction include the analysis of quantum gravity effects on the iron line shape in the reflection spectrum of the accretion disk around a RGI-Kerr BH \cite{r11}, the use of X-ray reflection spectroscopy of a Novikov–Page–Thorne (NPT) type disk in order to constraint the inverse dimensionless fixed-point parameter present in the RGI-Kerr metric \cite{r12}, the investigation of modifications to the shadow cast by RGI-Schwarzschild and RGI-Kerr BH \cite{r13,r14,r15,r16}, and the study of quantum gravity effects on radiation properties of a NPT accretion disk around a RGI-Schwarzschild BH \cite{r17}. For recent reviews on BH in asymptotically safe gravity see \cite{r18,r19}.\\

\noindent
In this work, we go a step forward and analyze quantum gravity modifications, in the infrared (IR) regime, to the thermal characteristics of a thin accretion disk spiraling around a RGI-Kerr BH, and confront our results with those for the classical Kerr BH of GR. To this purpose, we assume the relativistic NPT model for the thin disk \cite{r20,r21} and by varying the free parameter $\tilde{\xi}$ appearing in the RGI-metric and that encodes the AS quantum gravity effects, we calculate the modifications to the location of the innermost stable circular orbit (ISCO), the time averaged energy flux, the disk temperature, the observed luminosity, and the efficiency for conversion of rest mass energy of the material falling toward the BH into thermal energy.\\

\noindent
This paper is structured as follows. The IR-limit of the RGI-Kerr metric in the AS theory is discussed in In Sec.~\ref{sec:sec2}. The equations describing the equatorial geodesics for massive particles are obtained in Sec.~\ref{sec:sec3}. Sec.~\ref{sec:sec4} is devoted to the NPT thin accretion disk model and the equations describing its thermal properties. In Sec.~\ref{sec:sec5} we calculate the properties of the relativistic NPT accretion disks in the RGI-geometry and analyze the quantum gravity effects on its radiation properties by comparing with the GR description. In the last section, we conclude.
%
\section{\label{sec:sec2}Renormalization group improved Kerr metric}
The leading quantum gravity effects on the properties of black holes with nonzero angular momentum in the AS framework were discussed for the first time in \cite{r22} where the structure of the horizons, the ergosphere, the static limit surfaces as well as the phase space available for the Penrose process were discussed. Then, in \cite{r23}, was shown that in the infrared (IR) limit, the running of the Newton coupling with the energy scale takes the form
\begin{equation}\label{eq1}
G(r) = G_0\left(1-\frac{\xi}{r^2}\right),
\end{equation}
where $G_0$ is the known Newton constant and $\xi$ is a free parameter that determines the quantum effects on the spacetime geometry. Thus, after replacing Eq.~(\ref{eq1}) into the classical Kerr metric, the line element of the RGI-Kerr spacetime in Boyer-Lindquist coordinates and in units $c=G_{0}=1$, acquires the form
\begin{eqnarray}\nonumber
ds^2&=&-\Big(1-\frac{2M_\text{eff}(r) r}{\Sigma}\Big)dt^2-\frac{4aM_\text{eff}(r) r\sin^2\theta}{\Sigma}dtd\phi+\frac{\Sigma}{\Delta}dr^2+\Sigma{d\theta^2}\\\label{eq2}
&+&\sin^2\theta\bigg[r^2+a^2+\frac{2a^2 M_\text{eff}(r) r}{\Sigma}\sin^2\theta\bigg]d\phi^2,
\end{eqnarray}
where $a$ is the spin parameter and we have defined the ``effective" mass
\begin{equation}
M_\text{eff}(r) = M\Big(1-\frac{\xi}{r^2}\Big),
\label{eq3}
\end{equation}
and where
\begin{equation} 
\Delta=r^2-2Mr\left(1-\frac{\xi}{r^2}\right)+a^2,
\label{eq4}
\end{equation}
and
\begin{equation} 
\Sigma=r^2+a^2 \cos{\theta}.
\label{eq5}
\end{equation}
\noindent
Clearly, this Kerr-type metric reduces to the known Kerr metric in classical GR in the limit $\xi\rightarrow{0}$.\\

\begin{figure}
 \includegraphics[width=.40\linewidth]{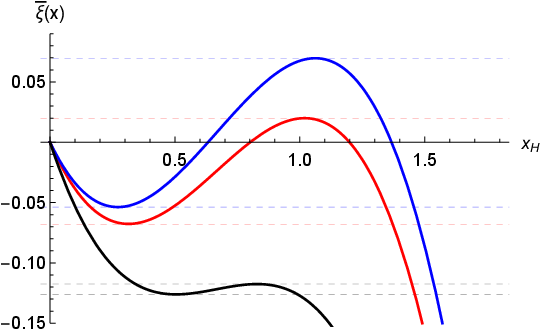}
 \includegraphics[width=.40\linewidth]{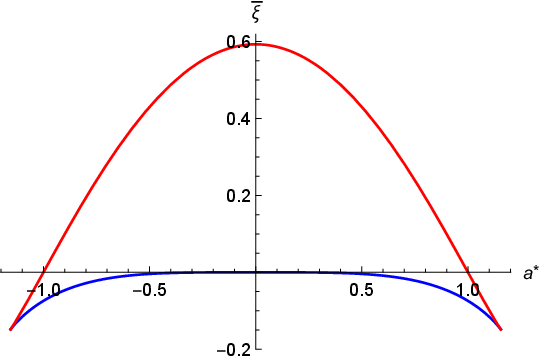}
\caption{Left panel: parameter $\tilde{\xi}$ as a function of the radial coordinate $x_H$ of the event horizon for $a^{\star}=0.93$ (blue), $a^{\star}=0.98$ (red), and for the non-classical allowed value $a^{\star}=1.12$ (black). The upper and lower horizontal dashed lines indicate the critical values $\tilde{\xi}_{c+}$ and $\tilde{\xi}_{c-}$, respectively, for each value of $a^{\star}$. Right panel: plots of the parameters $\tilde{\xi}_{c+}$ (red) and $\tilde{\xi}_{c-}$ (blue) as functions of the the dimensionless spin parameter $a^{\star}$.}
\label{F1}
\end{figure}

\noindent
The condition $\Delta=0$ determines the radii of the new horizons. This condition reads
\begin{equation}
x^3 - 2x^{2}+{a^{\star}}^{2} x+2\tilde{\xi}=0,
\label{eq6}
\end{equation}
where we have introduced the dimensionless parameters $x=r/M$, $\tilde{\xi}=\xi/M^{2}$ and $a^{\star}=a/M$. The discriminant of this cubic equation is
\begin{equation}
D_3=-{a^{\star}}^6+{a^{\star}}^4+(16-18 {a^{\star}}^2)\tilde{\xi}-27 {\tilde{\xi}}^2,
\label{eq7}
\end{equation}
For $D_3=0$, Eq.~(\ref{eq6}) with real coefficients has a multiple root and all the roots are reals. We first note that for $D_3=0$, $\tilde{\xi}=a^{\star}=0$ is solution to Eq.~(\ref{eq7}) which corresponds to $x=0$ in Eq.~(\ref{eq6}) (see Fig.~\ref{F1}). However, this solution has no physical meaning since the IR-approximation in Eq.~(\ref{eq1}) breaks around $x=0$. Now, the condition $D_3=0$ is a quadratic equation in $\tilde{\xi}$ with solutions
\begin{equation}
\tilde{\xi}_{c\pm} = \frac{\pm\sqrt{\left(4 -3 {a^{\star}}^{2}\right)^3}-9 {a^{\star}}^{2} + 8}{27},
\label{eq8}
\end{equation}
which implies that: (a) the allowed real values of $\tilde{\xi}$ require $4\geq 3 {a^{\star}}^{2}$ and are determined by the value of $a^{\star}$ in the range $\tilde{\xi}_{c-}\leq \tilde{\xi}<\tilde{\xi}_{c+}$, and (b) there exist a real solution for $a^{\star}=2/\sqrt{3}>1$ ($\tilde{\xi}=-4/27$) which has no classical analogue. This is illustrated in Fig.~\ref{F1} where, in the left panel we plot $\tilde{\xi}$ as a function of the radial coordinate of the event horizon $x_H$ for fixed values of $a^{\star}$, whereas in the right panel we plot $\tilde{\xi}_{c+}$ and $\tilde{\xi}_{c-}$ as functions of $a^{\star}$.  
\noindent
We see that the RGI-Kerr BH has two nonzero horizons only for $0\leq\tilde{\xi}<\tilde{\xi}_{c+}$, where $\tilde{\xi}_{c+}$ is the critical value for which the two horizons merge from below, such that for $\tilde{\xi}>\tilde{\xi}_{c+}$ a naked singularity develops. For $\tilde{\xi}_{c-}<\tilde{\xi}<0$, there are three nonzero horizons with $\tilde{\xi}_{c-}$ being the critical value for which two of the three nonzero horizons meet from above. Since we are interested in confronting our results with the ones for the classical Kerr solution which has two horizons, we will only consider values of $\tilde{\xi}$ in the range $\tilde{\xi}\in [0,\tilde{\xi}_{c+})$. It is important to remark that in this range the radius of the outer horizon decreases while the radius of the inner horizon increases as $\tilde{\xi}$ grows from $\tilde{\xi}=0$ to $\tilde{\xi}=\tilde{\xi}_{c+}$. For the discussion on the quantum gravity corrections to the ergosphere, to the static limit surfaces and to the ring singularity, we refer the reader to Refs.~\cite{r22,r23,r24}.\\

\section{\label{sec:sec3}Geodesics in the RGI-Kerr spacetime}
\begin{figure}
 \includegraphics[width=.48\linewidth]{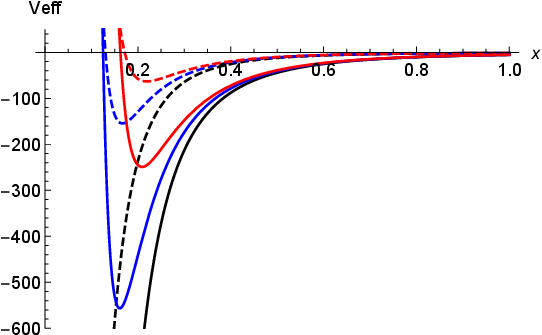}
\caption{Effective potential for $M=1$, $k=1$, $a^{\star}=0.5$, and for selected values of $h$ and $\tilde{\xi}$. Dashed lines correspond to prograde orbits ($h>0$) and solid lines are for retrograde motion ($h<0$). For dashed curves $h=2$ and: $\tilde{\xi}=0.025$ (red dashed), $\tilde{\xi}=0.015$ (blue dashed), and the classical case $\tilde{\xi}=0$ (black dashed). For solid lines $h=-2$ and: $\tilde{\xi}=0.025$ (red), $\tilde{\xi}=0.015$ (blue), and $\tilde{\xi}=0$ (black).}
\label{F2}
\end{figure}

By using the effective mass parameter, we can obtain the Lagrangian for the RGI-Kerr metric by simply replacing the mass $M$ of the BH by $M_{\rm{eff}}$ in the Lagrangian of the classical theory. In this way and in the equatorial plane $(\theta=\frac{\pi}{2},\dot{\theta}=0)$ we have
\begin{eqnarray}\nonumber
2L=&-&\left[1-\frac{2M_\text{eff}(r)}{r}\right]\dot{t}^2-\frac{4aM_\text{eff}(r)}{r}\dot{t}\dot{\phi}\\\label{eq9}
              &+&\frac{r^2}{\Delta}\dot{r}^2+\left[r^2+a^2+\frac{2a^2M_\text{eff}(r)}{r}\right]\dot{\phi}^2,
\end{eqnarray}
where the dots denote derivatives respect to the proper time $\tau$ that can be used as affine parameter for massive particles following timelike geodesics. Since $L$ does not depend on $t$ and $\phi$, the corresponding generalized momenta $p_t$ and $p_{\phi}$ are constant. Then, calling $k$ and $h$ the specific energy and angular momentum that is, the energy and angular momentum per unit rest mass of the particle following the orbit, from Eq.~(\ref{eq9}) we get
\begin{eqnarray}
p_t&=&-\left[1-\frac{2M_\text{eff}(r)}{r}\right]\dot{t}-\frac{2aM_\text{eff}(r)}{r}\dot{\phi}=-k,\label{eq10}\\
p_{r}&=&\frac{r^2}{\Delta}\dot{r}\label{eq11},\\
p_{\phi}&=&-\frac{2aM_\text{eff}(r)}{r}\dot{t}+\left[r^2+a^2+\frac{2a^2M_\text{eff}(r)}{r}\right]\dot{\phi}=h,\label{eq12}
\end{eqnarray}
The Hamiltonian, which is a constant as does not depend on $t$, acquires the form
\begin{eqnarray}\nonumber
2H&=&{-k}\dot{t}+h\dot{\phi}+\frac{r^2}{\Delta}\dot{r}\\\nonumber
&=&\left[-\left(1-\frac{2M_\text{eff}(r)}{r}\right)\dot{t}-\frac{2aM_\text{eff}(r)}{r}\dot{\phi}\right]\dot{t}\\\nonumber 
&+&\left[-\frac{2aM_\text{eff}(r)}{r}\dot{t}+\left(r^2+a^2+\frac{2a^2M_\text{eff}(r)}{r}\dot{\phi}\right)\right]\dot{\phi}+\frac{r^2}{\Delta}\dot{r}^2=C,\\\label{eq13}
\end{eqnarray}
where the constant $C$ takes the values $C=-1,0,1$ for timelike, null and spacelike geodesics, respectively. In this work we restrict ourselves to geodesics of massive particles, that is to the case $C=-1$. From Eqs.~(\ref{eq10}) and (\ref{eq12}) we get
\begin{eqnarray}
\dot{t}&=&\frac{1}{\Delta}\left[\left(r^2+a^2+\frac{2a^2M_\text{eff}(r)}{r}\right)k-\frac{2aM_\text{eff}(r)}{r}h\right],\label{eq14}\\
\dot{\phi}&=&\frac{1}{\Delta}\left[\frac{2aM_\text{eff}(r)}{r}k+\left(1-\frac{2M_\text{eff}(r)}{r}\right)h\right].\label{eq15}
\end{eqnarray}
Substituting Eqs.~(\ref{eq14}) and (\ref{eq15}) into Eq.~(\ref{eq13}), the radial equation of motion (also known as the ``energy" equation) is obtained as
\begin{eqnarray}
\frac{1}{2}\dot{r}^2+V_\text{eff}(r)=\frac{1}{2}\left(k^2-1\right),\label{eq16}
\end{eqnarray}
where
\begin{equation}\label{eq17}
V_\text{eff}(r)=-\frac{M_\text{eff}(r)}{r}+\frac{h^2-a^2(k^2-1)}{2r^2}-\frac{M_\text{eff}(r)(h-ak)^2}{r^3}
\end{equation}
is the effective potential per unit mass. Fig.~\ref{F2} are plots of the effective potential in terms of the dimensionless variables $x=r/M$, $\tilde{\xi}=\xi/M^2$ and $a^{\star}=a/M$. These plots show that, at difference with the classical solution, quantum gravity effects give rise to the presence of potential wells as a consequence of the fact that $V_{\text{eff}}(r)\rightarrow +\infty$ when $r \rightarrow 0$. It must be noted that the well of potential becomes shallower as the value of 
$\tilde{\xi}$ gets larger. 
%
\subsection{\label{sec:sec3.1}Circular orbits of massive particles}
In this section we follow the usual procedure (see, e.g., Ref.~\cite{r25}), as applied in Ref.~\cite{r23}, to calculate the circular geodesics of massive particles. The radial equation of motion, after the change of variable $u=1/r$, takes the form
\begin{eqnarray}\nonumber
F(u)=u^{-4}\dot{u}^2&=&k^2+2Mu^3\left(1-\xi u^2\right)\left(ak-h\right)^2+\left(a^2k^2-h^2\right)u^2\\\label{eq18}
&-&\left(1-2Mu\left(1-\xi u^2\right)+a^2u^2\right).
\end{eqnarray}
Writing $y=h-ak$, the conditions for the existence of circular orbits: $F(u)=0$ and $(dF/du)(u)=0$ produce, respectively
\begin{eqnarray}\label{eq19}
k^2-\left(1-2Mu\left(1-\xi u^2\right)+a^2 u^2\right)+2 M u^3\left(1-\xi u^2\right)y^2-\left(y^2+2yak\right)u^2=0,
\end{eqnarray}
\begin{eqnarray}\label{eq20}
M\left(1-3\xi u^2\right)-a^2 u+3 M u^2 y^2\left(1-\frac{5}{3}\xi u^2\right)-\left(y^2+2yak\right)u=0.
\end{eqnarray}
Solving these last two equations we have
\begin{eqnarray}\label{eq21}
k^2=Mu^3y^2\left(1-3\xi u^2\right)+1-Mu\left(1+\xi u^2\right),
\end{eqnarray}
\begin{eqnarray}\label{eq22}
2yaku=3M u^2 y^2\left(1-\frac{5}{3}\xi u^2\right)+M\left(1-3\xi u^2\right)-y^2 u-a^2 u.
\end{eqnarray}
Using Eqs.~(\ref{eq21}) and (\ref{eq22}) to eliminate $k$, the following quadratic equation in $y^2$ is obtained
\begin{eqnarray}\nonumber
&y^4&u^2\left[\left(3Mu\left(1-\frac{5}{3}\xi u^2\right)-1\right)^2-4M a^2 u^3\left(1-3\xi u^2\right)\right]\\\nonumber
&-&2y^2 u\left[\Big(3Mu\left(1-\frac{5}{3}\xi u^2\right)-1\Big)\Big(a^2u-M(1-3\xi u^2)\Big)+2a^2u(1-Mu(1+\xi u^2))\right]\\\label{eq23}
&+&\Bigg[a^2 u-M(1-3\xi u^2)\Bigg]^2=0.
\end{eqnarray}
The discriminant of this equation is given by
\begin{eqnarray}\label{eq24}
D_2=16a^2Mu^3(1-3\xi u^2)\left({\Delta_u}\right)^2,
\end{eqnarray}
where $\Delta_u=1+a^2u^2-2Mu\left(1-\xi u^2\right)$. The calculation of the solutions to Eq.~(\ref{eq23}) is eased by writing
\begin{eqnarray}\label{eq25}
\left(1-3Mu\left(1-\frac{5}{3}\xi u^2\right)\right)^2-4M a^2 u^3\left(1-3\xi u^2\right)=F_+F_-,
\end{eqnarray}
where
\begin{eqnarray}\label{eq26}
F_\pm=1-3Mu\left(1-\frac{5}{3}\xi u^2\right)\pm{2au\sqrt{Mu\left(1-3\xi u^2\right)}}.
\end{eqnarray}
Replacing into Eq.~(\ref{eq23}), we find
\begin{eqnarray}\label{eq27}
y^2u^2=\frac{\Delta_u F_{\pm}-F_+F_-}{F_+F_-}=\frac{\Delta_u-F_{\mp}}{F_{\mp}}=\frac{u\left[a\sqrt{u}\pm\sqrt{M\left(1-3\xi u^2\right)}\right]^2}{F_{\mp}},
\end{eqnarray}
In this way, for stable circular orbits, we obtain the following solutions for $y$
\begin{eqnarray}\label{eq28}
y=-\frac{a\sqrt{u}\pm\sqrt{M\left(1-3\xi u^2\right)}}{\sqrt{u F_\mp}}.
\end{eqnarray}
\noindent
Using Eqs.~(\ref{eq28}), (\ref{eq21}) and the definition $y=h-ak$, we get the specific energy and the specific angular moment of circular orbits. In terms of the dimensionless variables $x$, $\tilde{\xi}$ and $a^{\star}$, they are:
\begin{eqnarray}\label{eq29}
k=\frac{x^3-2(x^2-\tilde{\xi})\mp a^{\star}\sqrt{x\left(x^2-3\tilde{\xi}\right)}}{x^{3/2}\sqrt{x^3-3x^2+5\tilde{\xi} \mp 2a^{\star} \sqrt{x\left(x^2-3\tilde{\xi} \right)}}},
\end{eqnarray}
\begin{eqnarray}\label{eq30}
h=\mp M \frac{(x^2+a^{\star 2})\sqrt{x\left(x^2-3\tilde{\xi} \right)}\pm 2a^{\star}(x^2-\tilde{\xi})}{x^{3/2}\sqrt{x^3-3x^2+5\tilde{\xi} \mp 2a^{\star} \sqrt{x\left(x^2-3\tilde{\xi} \right)}}},
\end{eqnarray}
where upper and lower signs correspond to counter-rotating (retrograde) and co-rotating (prograde) orbits, respectively.\\

\begin{figure}
\centering
    \includegraphics[width=0.40\textwidth]{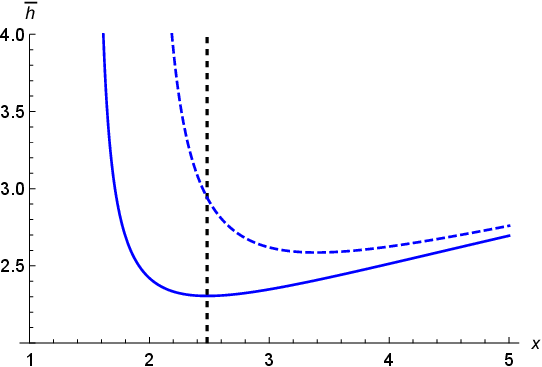}
    \includegraphics[width=0.40\textwidth]{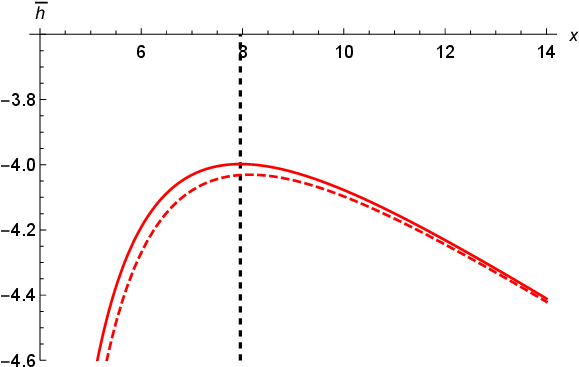}
\caption{The angular momentum $\tilde{h}$ as a function of $x$ for co-rotating motion (left) and for counter-rotating motion (right) when $a^{\star}=0.7$ for different values of $\tilde{\xi}$ . In the left panel $\tilde{\xi=0.25}$ (blue solid line) and $\tilde{\xi}=0$ (blue dashed line). The same values of $\tilde{\xi}$ for the right panel. The vertical dashed lines are the position of $x_{\rm isco}$ for both cases.
   }\label{F3}
\end{figure}

\noindent
From Eqs.~(\ref{eq14}) and (\ref{eq15}) we calculate the angular velocity $\Omega$ which acquires the form
\begin{eqnarray}\label{eq31}
\Omega=\frac{\dot{\phi}}{\dot{t}}=\mp\frac{\left(x^2-3\tilde{\xi} \right)^{1/2}}{x^{5/2}\mp a^{\star}\left(x^2-3\tilde{\xi} \right)^{1/2}}.
\end{eqnarray}
Now, the radius of the innermost stable circular orbit $x_\text{isco}$ occurs at the local minimum of the effective potential, so is calculated from
\begin{eqnarray}\label{eq32}
\frac{d^2V_\text{eff}}{dx^2}|_{x=x_\text{isco}}=0.
\end{eqnarray}
The explicit form of this equation is not very illuminating. To illustrate the main results, in Table~\ref{tab:1} we have listed the value of $x_{\rm isco}$ for different values of $\tilde{\xi}$ both for co-rotating and for counter-rotating motion and for a nearly extreme Kerr BH with $a^{\star}=0.98$ for which $\tilde{\xi}_{c+}=0.0199$. In Table~\ref{tab:2}, we do the same for a slow rotating BH with $a^{\star}=0.3$ and $\tilde{\xi}_{c+}=0.5331$. From these tables it follows that, for all the cases, as 
$\tilde{\xi}$ grows, the radius of the ISCO goes to smaller values, and that for a high spin BH, this reduction impact the prograde motion more than the retrograde one. The same happens for low spin BHs but the difference in the decreasing of the ISCO is not so prominent. This shouldn't be surprising because, as is well known, retrograde orbits are easily destabilized by frame dragging, shifting the ISCO away from the black hole. This means that for retrograde geodesics the Lense-Thirring effect is operative in counteracting the quantum effect of reduction of the ISCO radius. Fig.~\ref{F3} shows plots of the angular momentum $\tilde{h}=h/M$ vs $x$ for prograde (left) and for retrograde motion (right). We see that for $\tilde{\xi}\neq 0$ and in the vicinity of the ISCO, the values of the angular momentum start to be displaced toward smaller values and that, at $x_{\rm isco}$, this quantum gravity effect is about $10$ times greater for co-rotating than for counter-rotating circulation, as compared with the classical values.\\

\begin{figure}
\includegraphics[width=.48\linewidth]{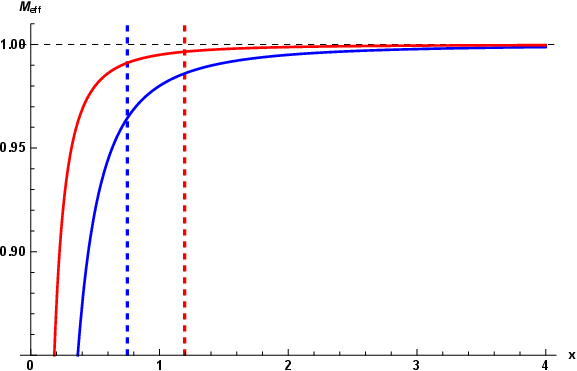}
\caption{Effective mass versus $x$ for $M=1$ and for $\tilde{\xi}=0.02$ (blue) and $\tilde{\xi}=0.005$ (red). The vertical dashed lines locate the ISCO for the same values of $\tilde{\xi}$, respectively.}
\label{F4}
\end{figure}

\noindent
The behavior of $x_{\rm isco}$ and $\tilde{h}$ with growing values of $\tilde{\xi}$ seems to be in conflict with the antiscreening nature of gravity at high energies according the AS program. However, these results in the IR-limit have a simple physical interpretation \cite{r17}: notwithstanding the angular momentum decreases with increasing values of $\tilde{\xi}$ both for prograde and retrograde motion, the effective mass $M_{\rm{eff}}$ decreases (for a fixed value of $\tilde{\xi}$) with decreasing values of $x$ (see Fig.~\ref{F4}) and, consequently, a particle falling into the BH feels a gravitational pull that weakens as it is accreted. This, in turn, makes the particle can stay in a stable circular geodesic with a smaller ISCO radius.\\
\begin{table}[!ht]
\centering
 \begin{ruledtabular}
 \begin{tabular}{p{4cm}p{4cm}|p{4cm}p{4cm}}
 \multicolumn{2}{c}{Co-rotating ($a^{\star}=0.98$, $\tilde{\xi}_{c+}=0.0199$)} & \multicolumn{2}{c}{Counter-rotating ($a^{\star}=0.98$, $\tilde{\xi}_{c+}=0.0199$)}\\ \hline
 $\tilde{\xi}$ & $x_{\rm isco}$ & $\tilde{\xi}$ & $x_{\rm isco}$\\ \hline
 0 & 1.6140 & 0 & 8.9437\\
 0.010 & 1.4715 & 0.010 & 8.9373\\
 0.019 & 1.2075 & 0.019 & 8.9315\\
 \end{tabular}
 \end{ruledtabular}
\caption{\label{tab:1}The value of $x_{\rm isco}$ for $a^{\star}=0.98$ and for different values of $\tilde{\xi}$. The left column is for prograde motion while the right column is for retrograde motion.}
\end{table}
\begin{table}[!ht]
\centering
 \begin{ruledtabular}
 \begin{tabular}{p{4cm}p{4cm}|p{4cm}p{4cm}}
 \multicolumn{2}{c}{Co-rotating ($a^{\star}=0.3$, $\tilde{\xi}_{c+}=0.5331$)} & \multicolumn{2}{c}{Counter-rotating ($a^{\star}=0.3$, $\tilde{\xi}_{c+}=0.5331$)}\\ \hline
 $\tilde{\xi}$ & $x_{\rm isco}$ & $\tilde{\xi}$ & $x_{\rm isco}$\\ \hline
 0 & 4.9786 & 0 & 6.9493\\
 0.40 & 4.3171 & 0.25 & 6.7205\\
 0.50 & 4.0861 & 0.50 & 6.4659\\
 \end{tabular}
 \end{ruledtabular}
\caption{\label{tab:2}The value of $x_{\rm isco}$ for $a^{\star}=0.3$ and for different values of $\tilde{\xi}$. The left column is for prograde motion while the right column is for retrograde motion.}
\end{table}

\section{\label{sec:sec4}The Relativistic Novikov-Page-Thorne model of the accretion disk}
The Novikov-Page-Thorne (NPG) model \cite{r20,r21} of the accretion disk around a BH is the relativistic generalization of the non-relativistic model of Shakura-Sunyaev \cite{r26}. The model assumes that: (a) the disk lies in the equatorial plane of the rotating BH and is thin, which means that $H/R << 1$ being $H$ the maximum half-thickness of the disk and $R$ its radius, (b) the particles in the disk follow nearly circular orbits with a small inward motion due to viscous torques. This results in an outward transportation of angular momentum. The loss of potential energy associated to the inward motion heat up the disk and the thermal energy is efficiently radiated away, (c) the disk is assumed in thermodynamic equilibrium in a quasi-steady state and the quantities describing its thermal properties are averaged over the azimuthal angle $\phi=2\pi$, over the height $H$, and over the characteristic time scale for a period of the orbits, and (d) the averaged radiation properties of the disk are calculated from the laws of conservation of rest mass, energy and angular momentum. In this way, the flux of radiant energy is obtained as
\begin{equation}\label{eq33}
{\cal F}(r)=-\frac{\dot{M}}{4\pi\sqrt{-g_{I}}}\frac{1}{(k-\Omega h)^2}\frac{d\Omega}{dr} \int_{r_{\rm isco}}^{r}(k-\Omega h) \frac{dh}{dr}dr,
\end{equation}
where $\dot{M}$ is the mass accretion rate which we assume as a constant, and $-g_{I}=r^2$ is the metric determinant in the equatorial plane of the RGI-BH. 
\noindent
From the assumed thermal equilibrium, the emitted radiation is black body with the temperature given by the Stefan-Boltzmann law
\begin{equation}\label{eq34}
T(r)=\sigma^{-1/4} {\cal F}(r)^{1/4},
\end{equation}
The observed luminosity $L\left(\nu \right)$, having a redshifted black body spectrum, is given by \cite{r27}
\begin{equation}
L\left(\nu \right) =4\pi d^{2}I\left(\nu \right) =\frac{8\pi h \cos i}{ c^2 } \int_{r_{i}}^{r_{f}}\int_0^{2\pi}\frac{\nu^{3}_e r d\phi dr }{\exp \left(\nu_e/T\right) -1},\label{eq35}
\end{equation}
where $I_{\nu}$ is the Planck distribution function, $d$ is the distance to the source, $i$ is the disk inclination angle, while $r_{i}$ and $r_{f}$ denote the positions of the inner and outer edges of the disk, respectively. The frequency of the emitted radiation is $\nu_e=\nu(1+z)$, with the redshift factor $z$ given by \cite{r28,r29}
\begin{equation}\label{eq36}
1+z=\frac{1+\Omega r \sin \phi \sin i }{\sqrt{ -g_{tt} - 2 \Omega g_{t\phi} - \Omega^2 g_{\phi\phi}}},
\end{equation}
\noindent
where $g_{\alpha \beta}$ are the metric coefficients (see Eq.~(\ref{eq2})) and $\Omega$ is the angular velocity given by Eq.~(\ref{eq31}).\\

\noindent
The efficiency $\epsilon$ for conversion of rest mass energy into radiant energy is defined as the energy lost by a particle that goes from the infinity, where $k=1$, to the inner boundary of the disk. Then, assuming that the totality of the emitted photons escape to infinity, we have
\begin{equation}\label{eq37}
\epsilon=\frac{k_{\infty}-k_{\rm isco}}{k_{\infty}}\approx 1 - k_{\rm isco}.
\end{equation}
%
\section{\label{sec:sec5}Radiation from a thin accretion disk around a RGI-Kerr Black Hole}
In this section we analyze deviations, respect to the classical GR predictions, arising from quantum gravity effects on the thermal properties (energy flux, temperature and observed luminosity) of a thin disk accreted by a RGI-Kerr BH in the IR-limit. Both for illustrative and comparison purposes, we fix the following relevant parameters: (a) the mass of the BH is fixed at the value $M=1$, (b) we take as a constant provided by the observations the rate at which material is removed onto the BH (the mass accretion rate $\dot{M}$), (c) we put $i=40^\circ$ for the disk inclination angle, and (d) we take $x_i=x_{\rm isco}$ as the inner edge of the disk and $x_f=30 x_i$ as the outer edge. We remark that, for our purposes of comparison with the predictions of GR, our results are independent of the location of the outer edge.\\

\noindent
\begin{figure}[t]
    \includegraphics[width=0.40\linewidth]{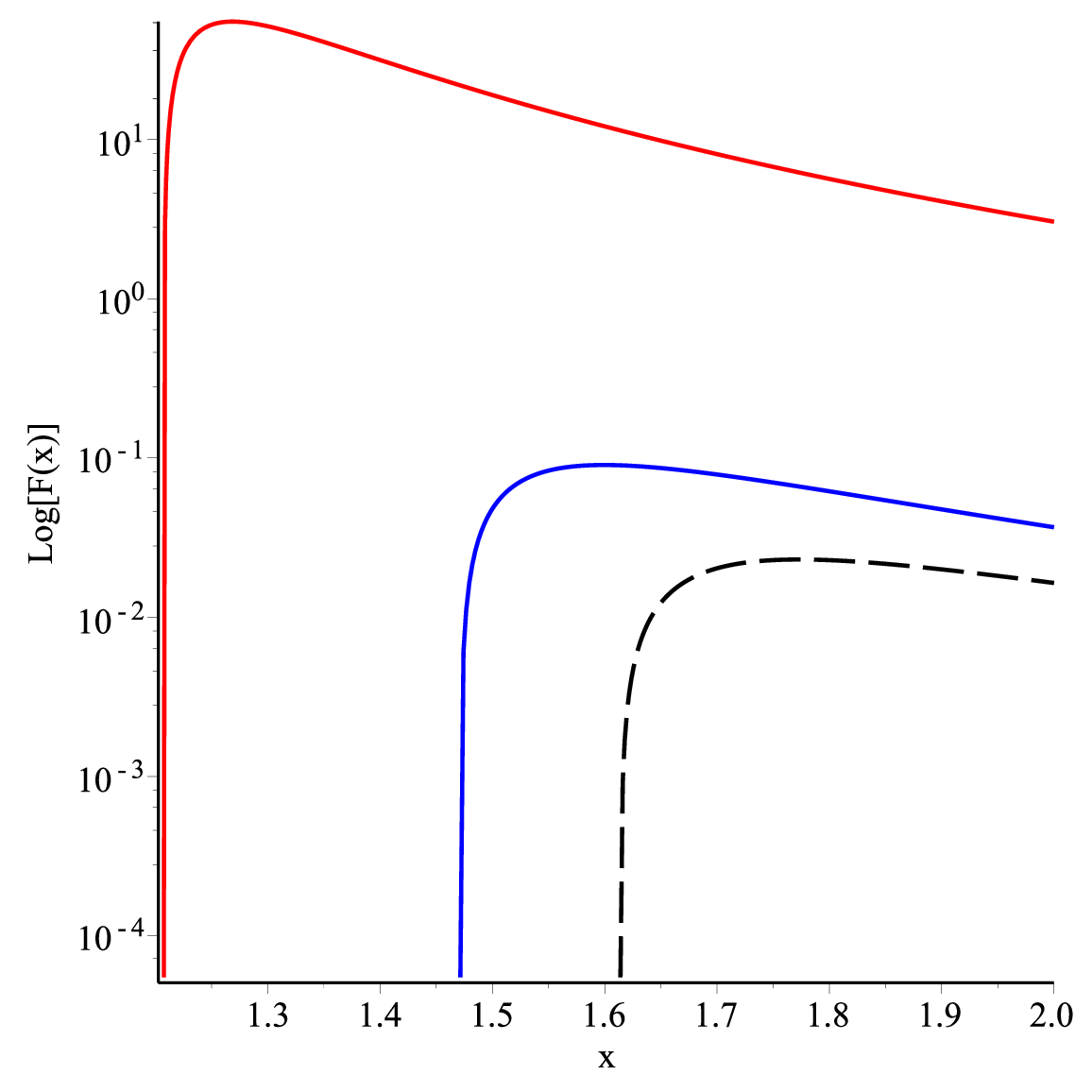}
    \includegraphics[width=0.40\linewidth]{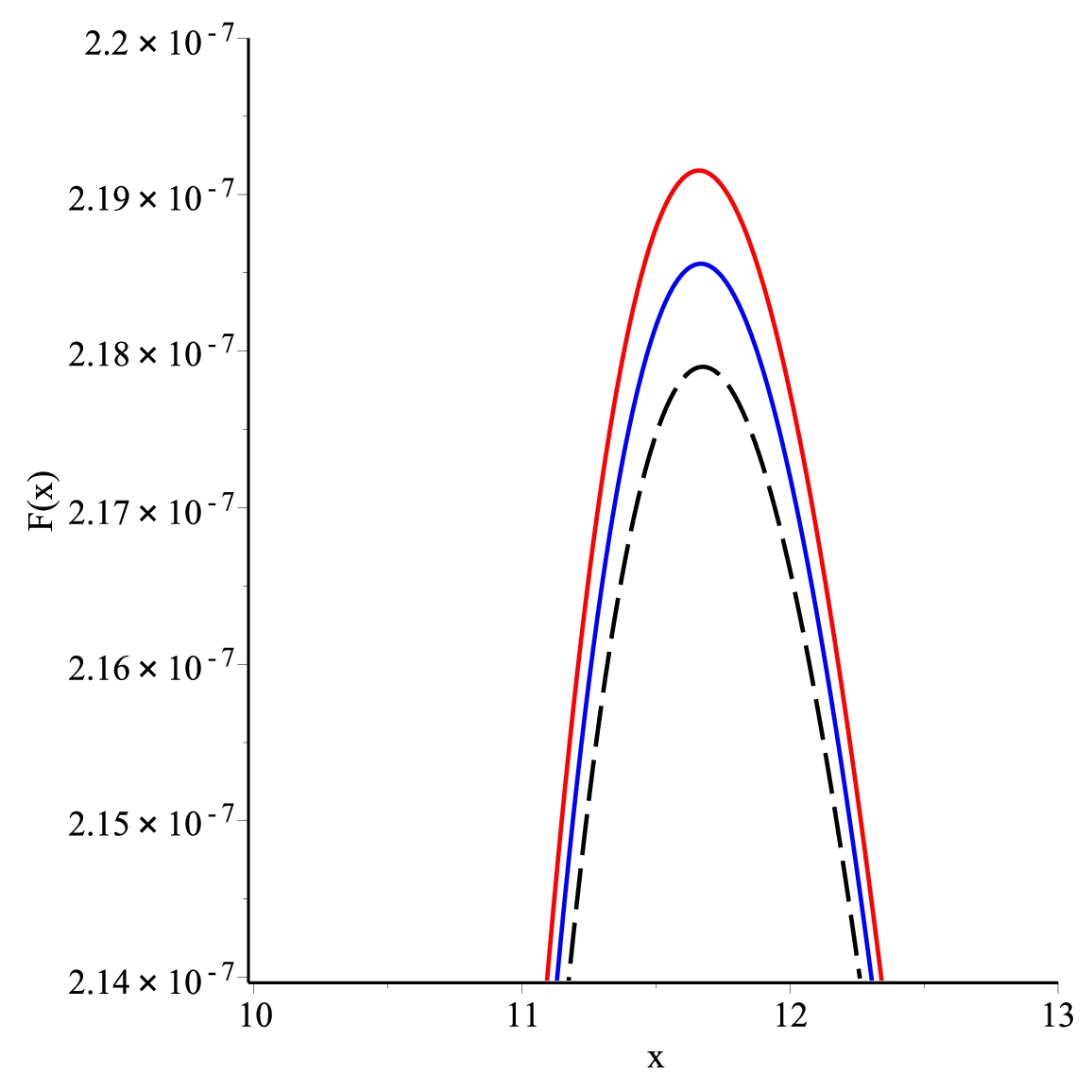}
     \includegraphics[width=0.40\linewidth]{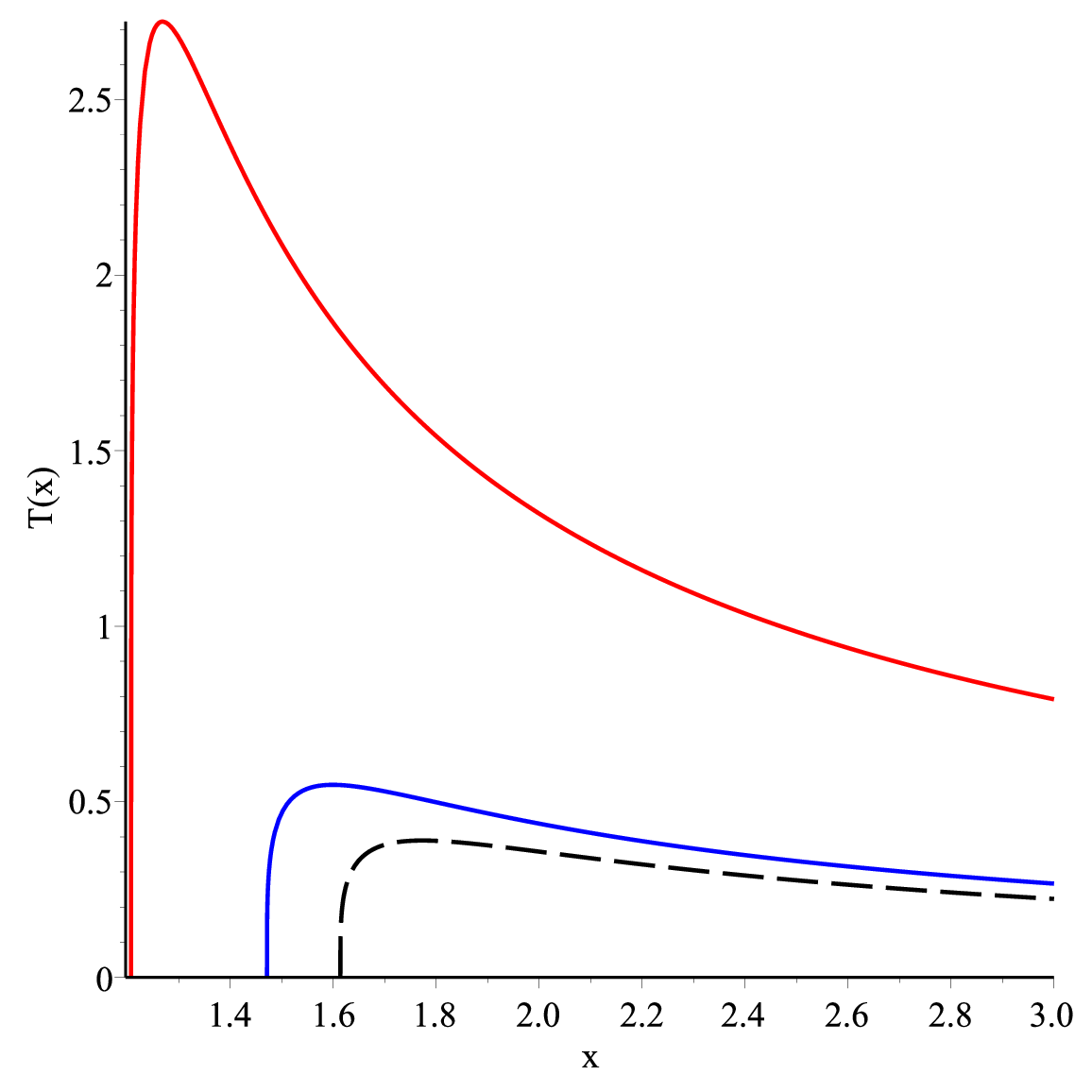}
     \includegraphics[width=0.40\linewidth]{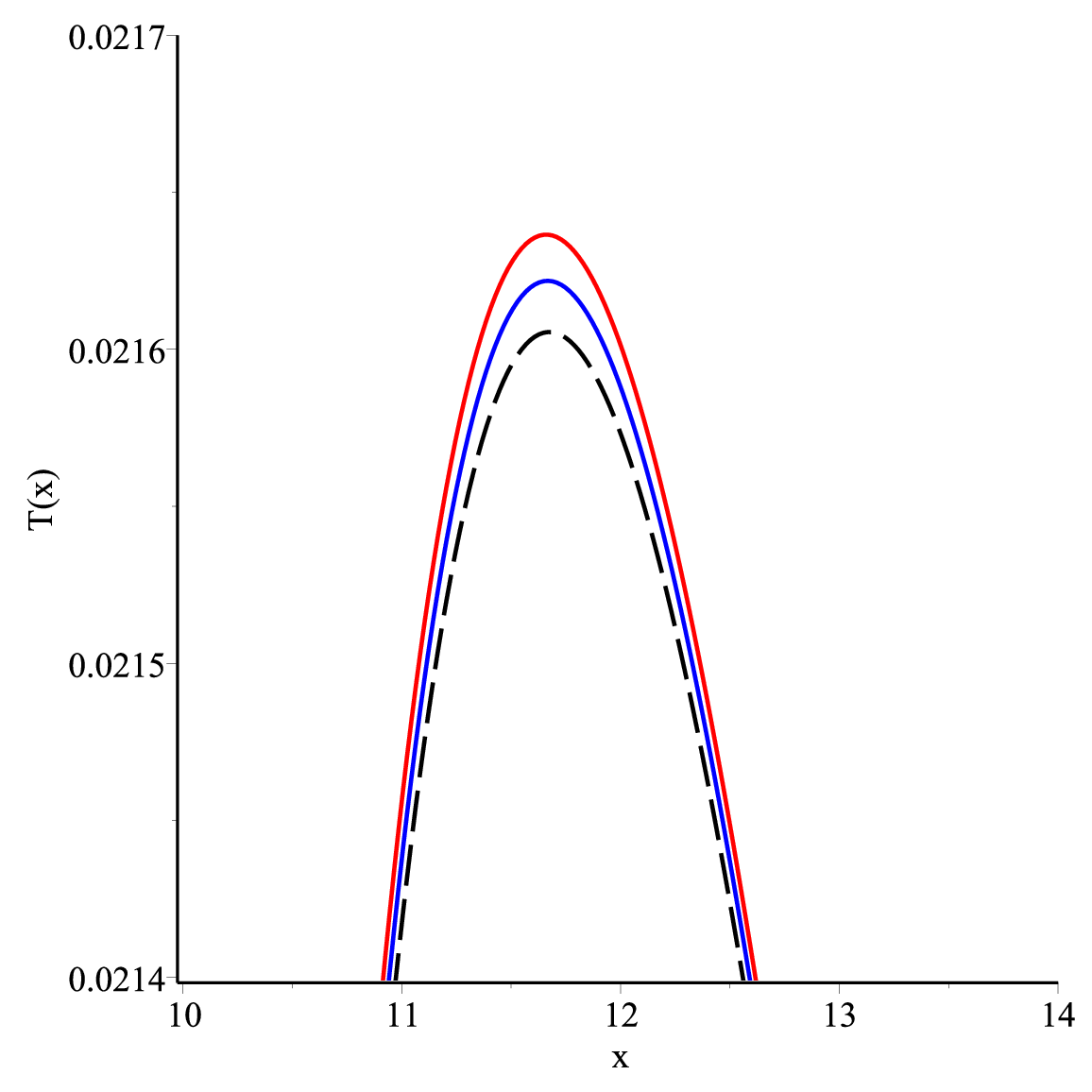}
      \includegraphics[width=0.40\linewidth]{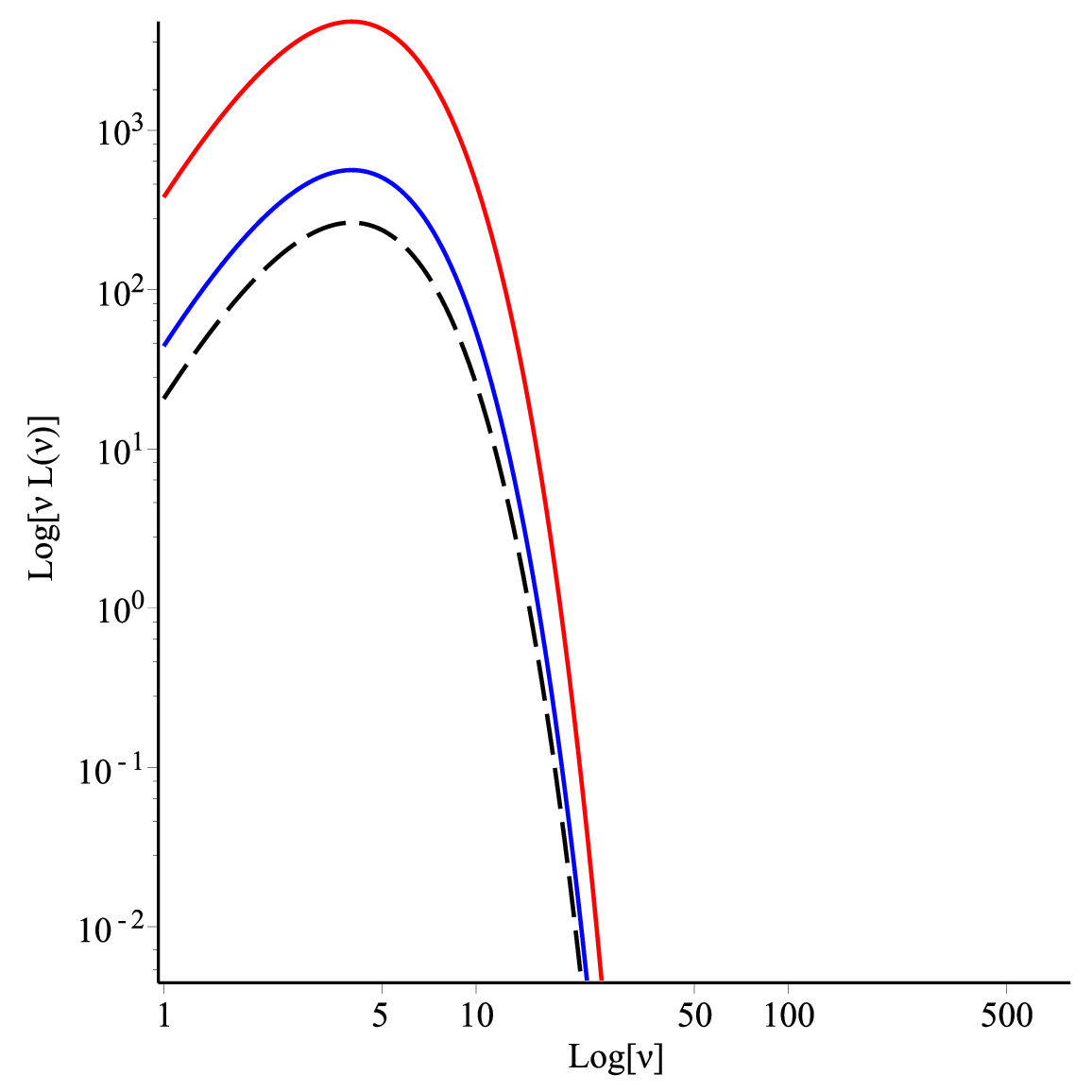}
      \includegraphics[width=0.40\linewidth]{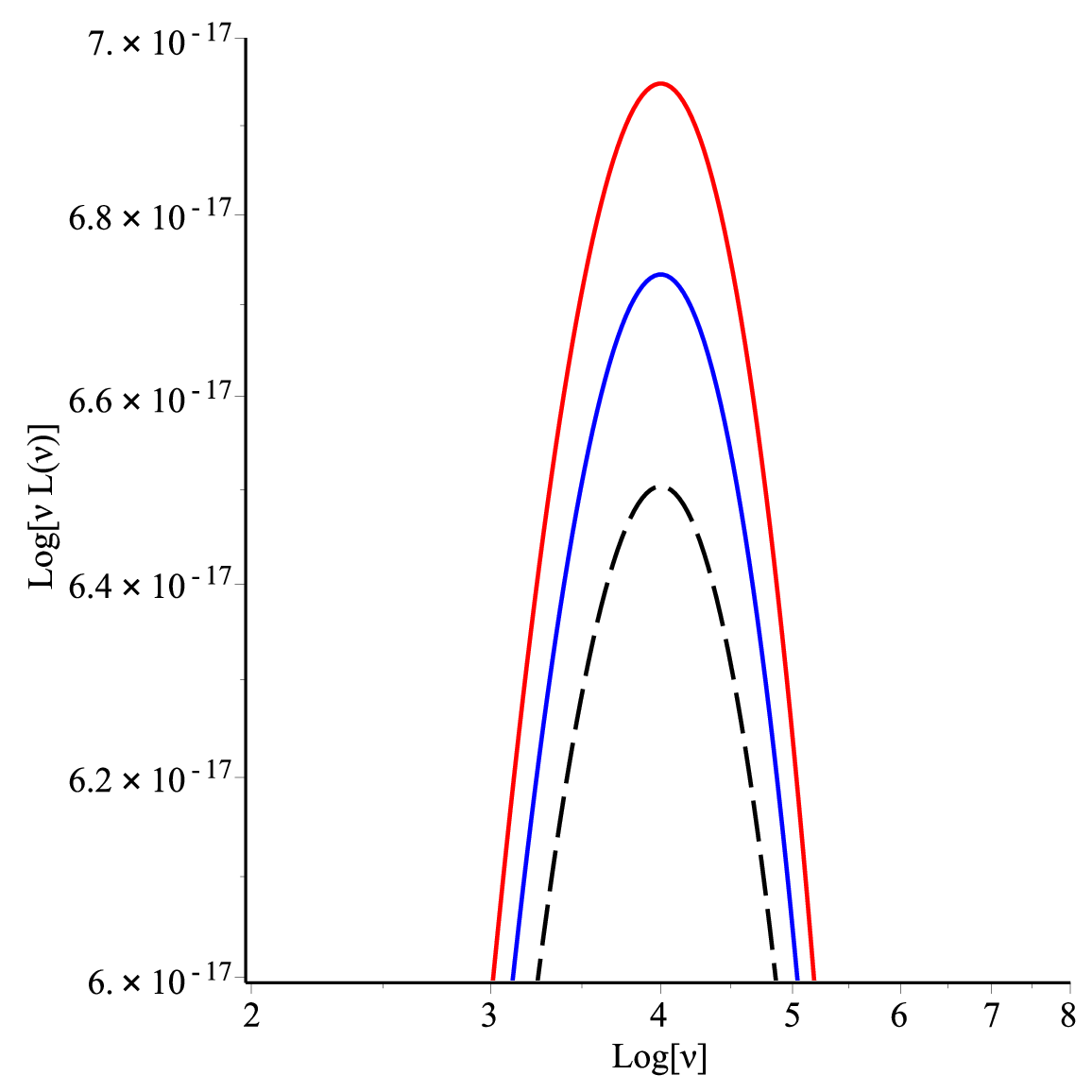}
\caption{Left from top to bottom: Plots of the energy flux, the temperature and the observed luminosity from a rapid rotating ($a=0.98$) RGI-Kerr BH with a prograde thin accretion disk for $\tilde{\xi}=0.019$ (red), $\tilde{\xi}=0.010$ (blue), and for the classical Kerr BH: $\tilde{\xi}=0$ (black dashed). Right: The same as for the left plots but for retrograde rotation.} \label{F5}
\end{figure}
\noindent
\begin{figure}[t]
    \includegraphics[width=0.40\linewidth]{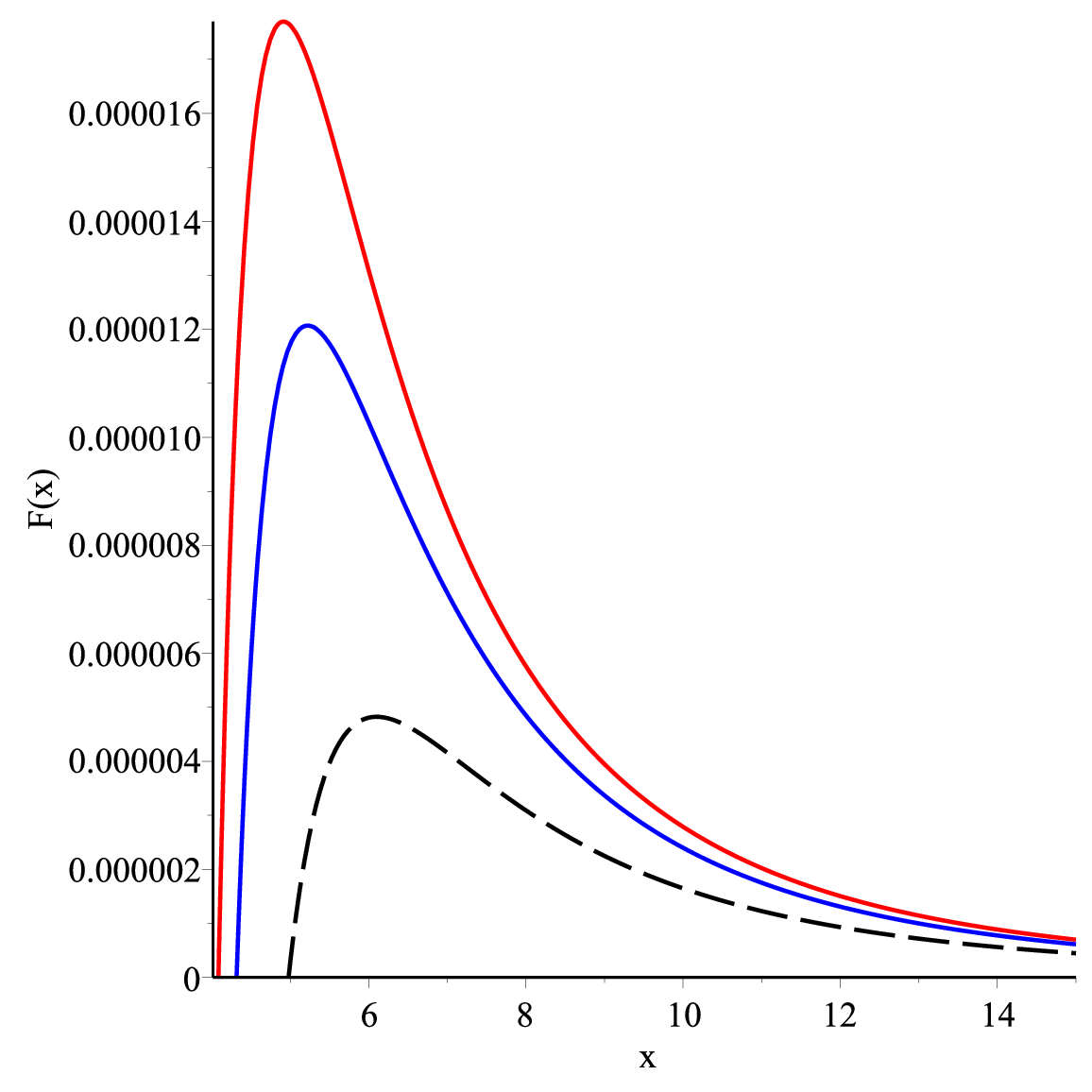}
    \includegraphics[width=0.40\linewidth]{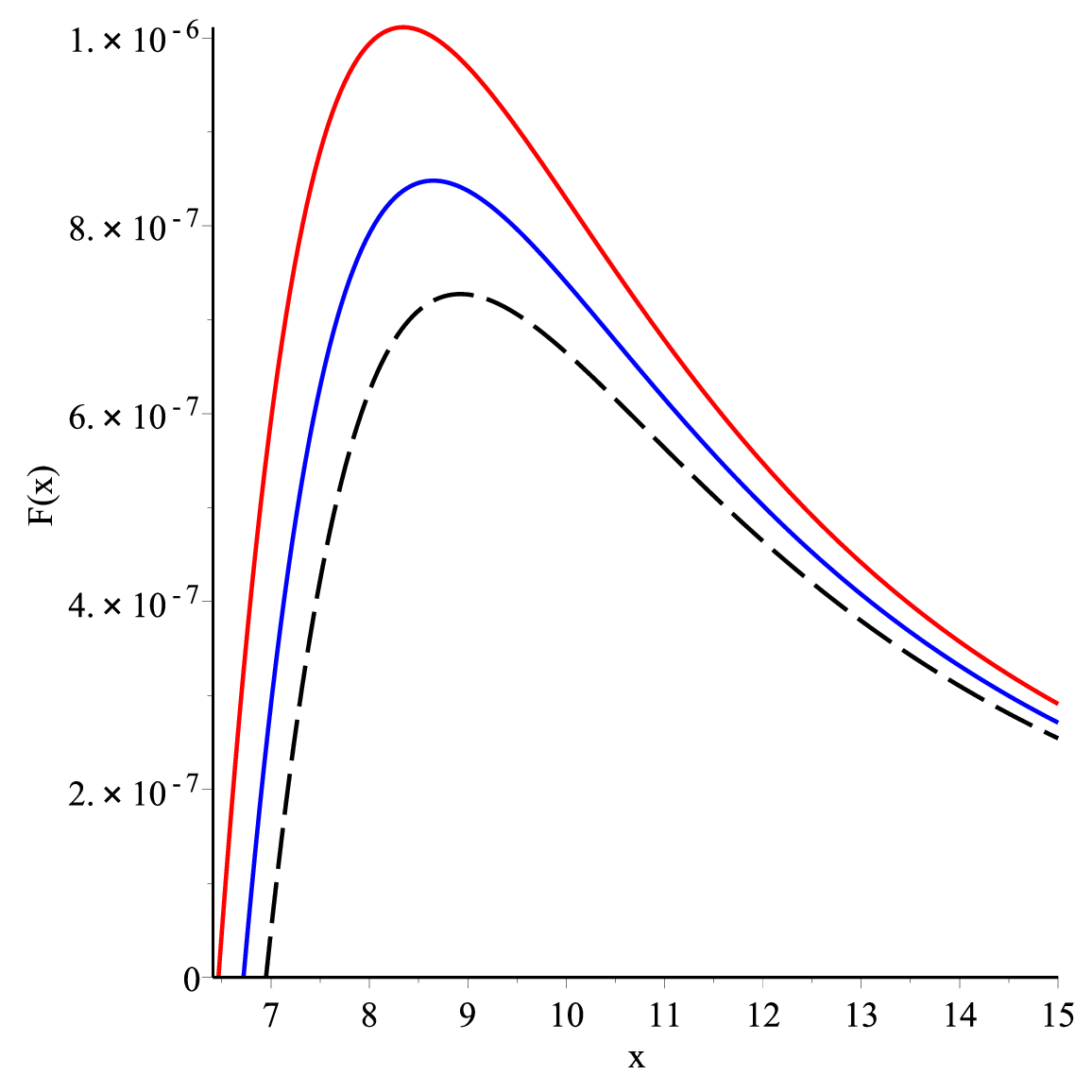}
     \includegraphics[width=0.40\linewidth]{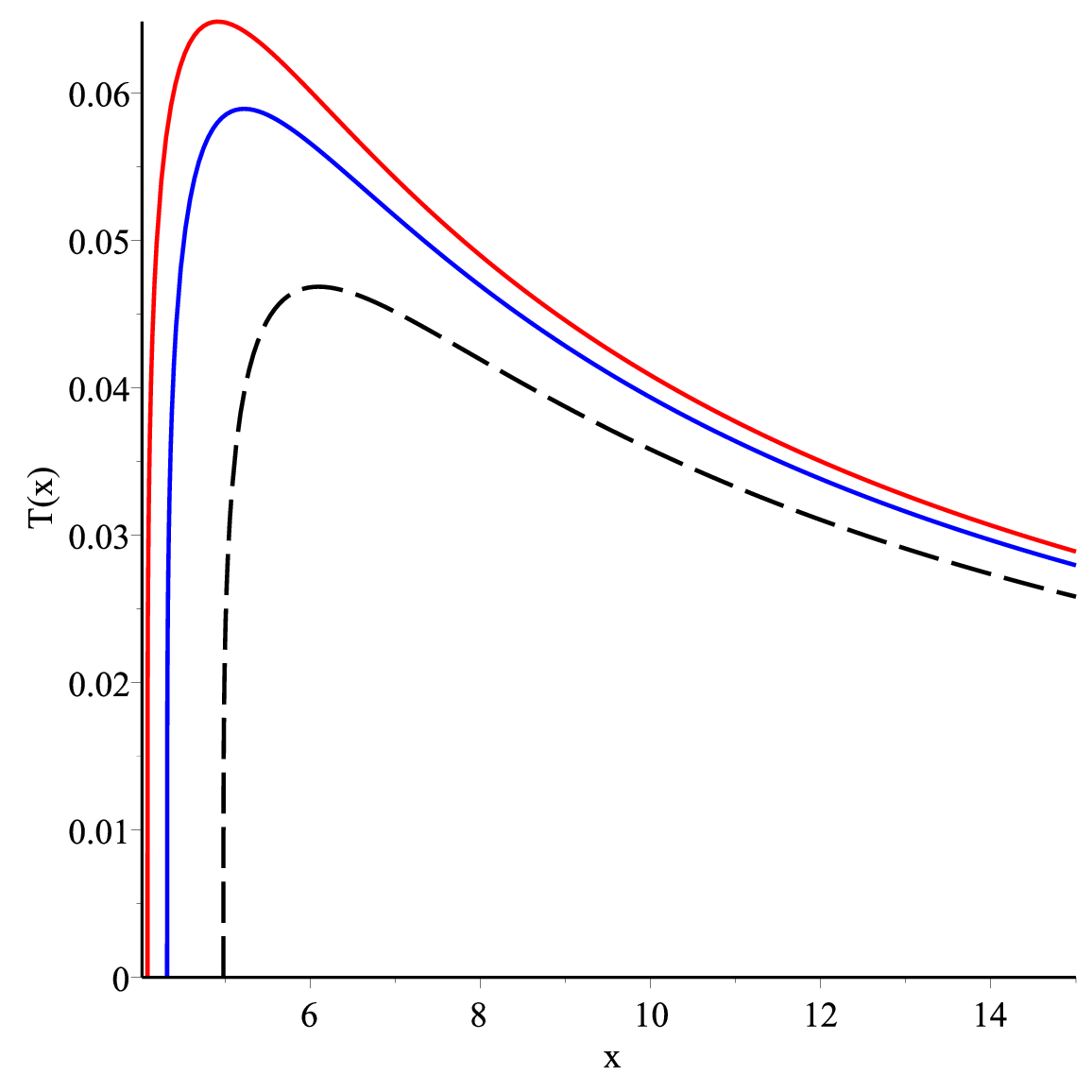}
     \includegraphics[width=0.40\linewidth]{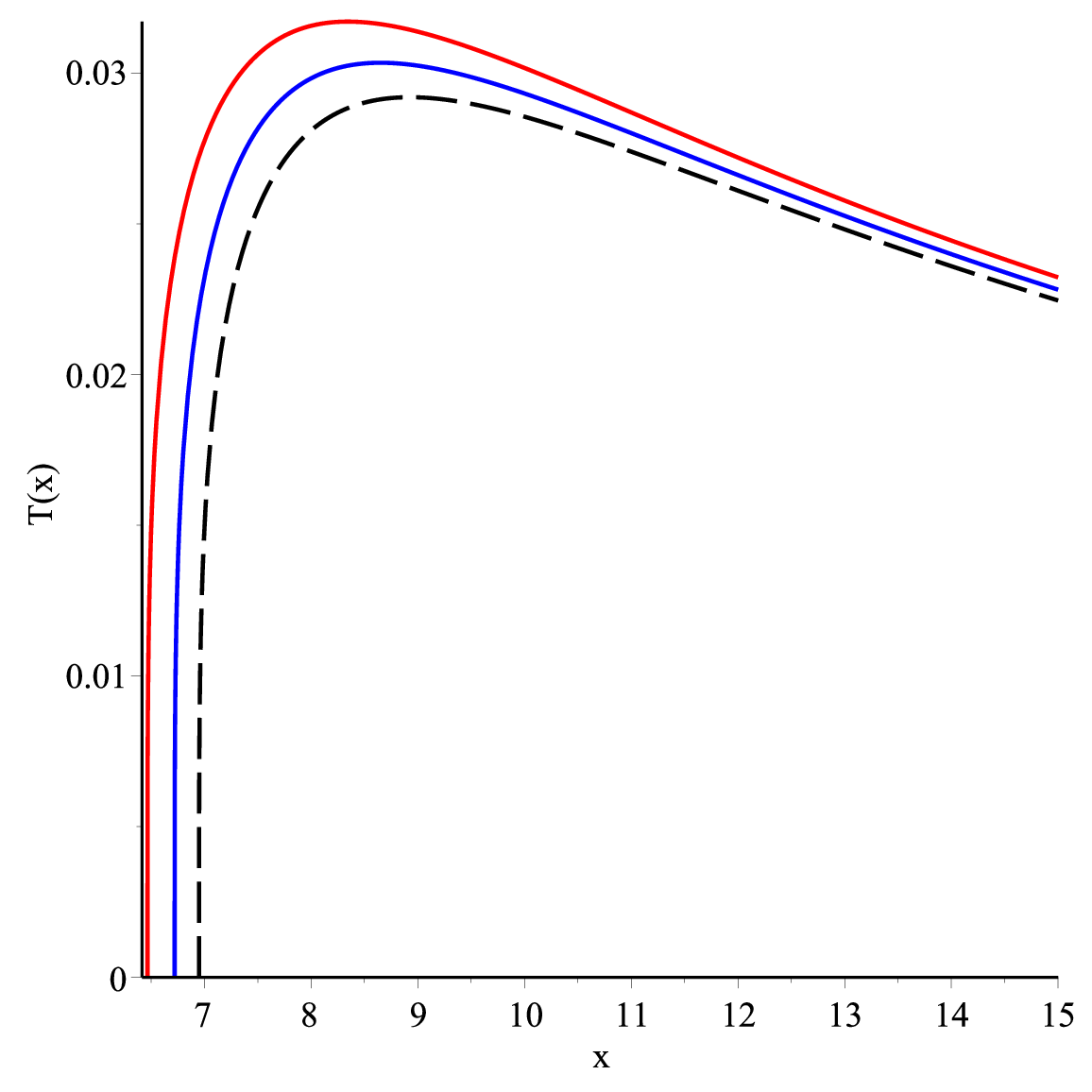}
      \includegraphics[width=0.40\linewidth]{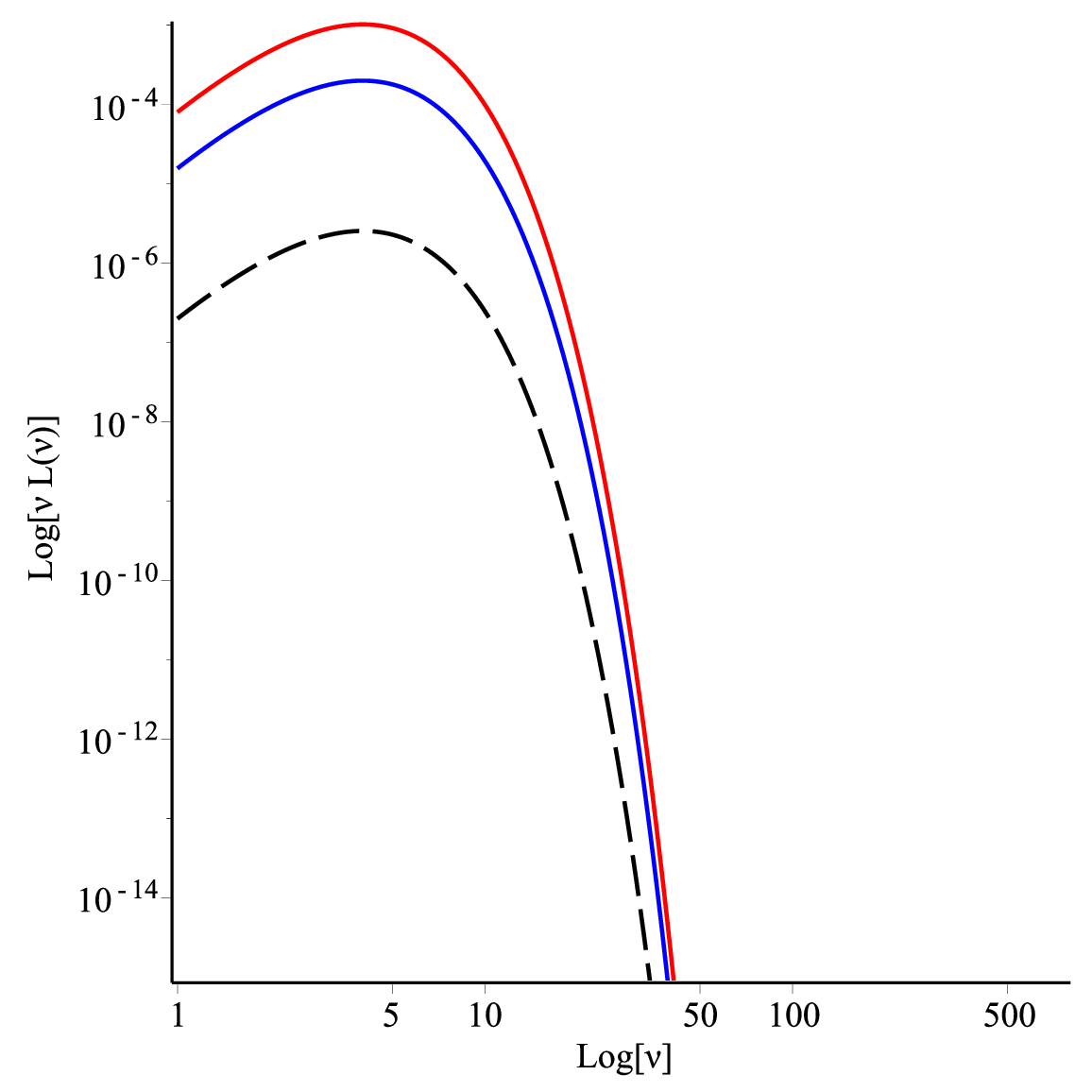}
      \includegraphics[width=0.40\linewidth]{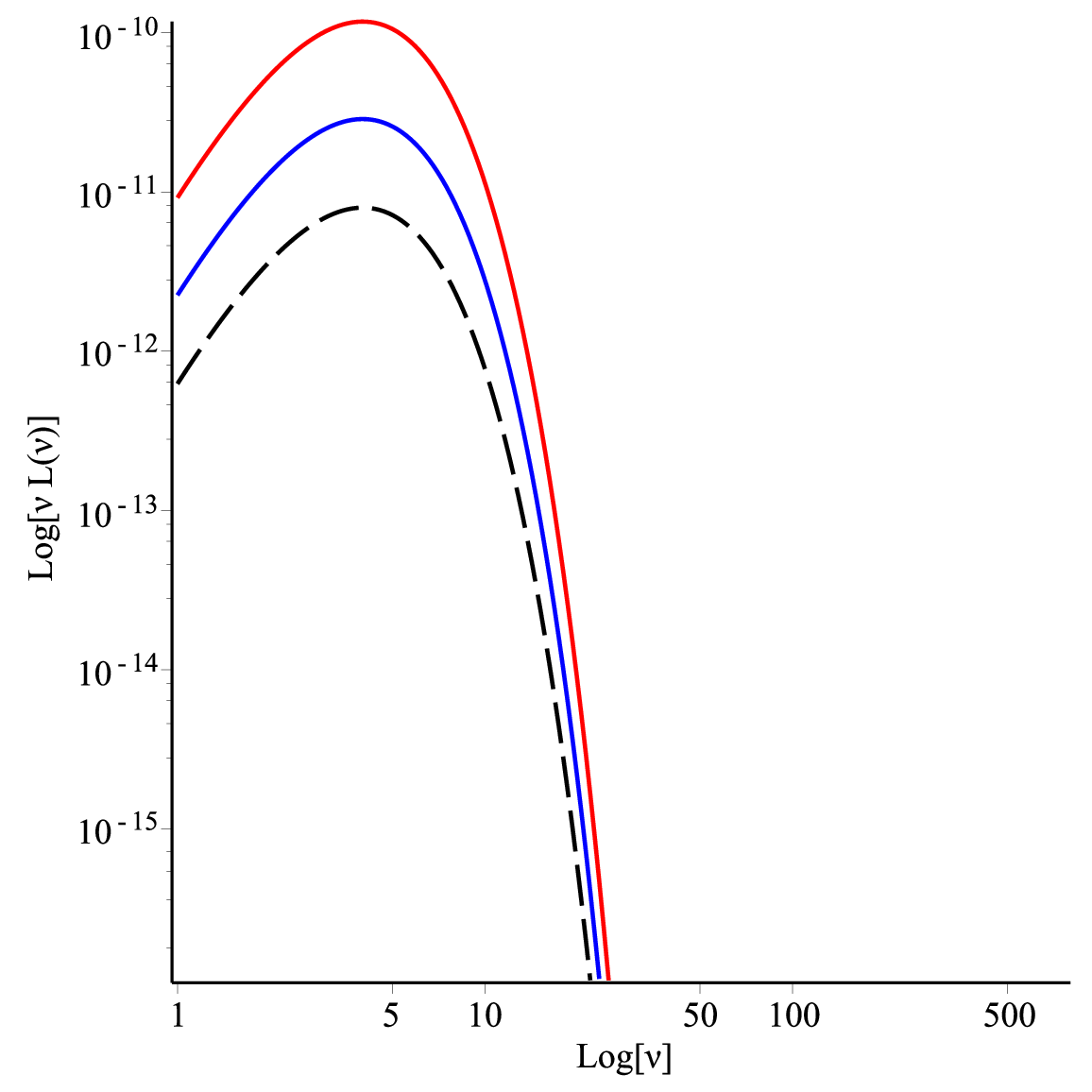}
\caption{Left from top to bottom: Plots of the energy flux, the temperature and the observed luminosity from a slow rotating ($a=0.3$) RGI-Kerr BH with a prograde thin accretion disk for $\tilde{\xi}=0.50$ (red), $\tilde{\xi}=0.40$ (blue), and for the classical Kerr BH: $\tilde{\xi}=0$ (black dashed). Right: The same as for the left plots but for retrograde rotation and for $\tilde{\xi}=0.5$ (red), $\tilde{\xi}=0.25$ (blue), and $\tilde{\xi}=0$ (black dashed).} \label{F6}
\end{figure}
\noindent
Figs.~\ref{F5} and \ref{F6} show the behavior of the above referred radiation properties using the values of $a^{\star}$, 
$\tilde{\xi}$ and $x_{\rm isco}$ recorded in Tables~\ref{tab:1} and \ref{tab:2}. Fig.~\ref{F5} is for a rapid rotating BH with $a^{\star}=0.98$, and Fig.~\ref{F6} is for a low spin BH with $a^{\star}=0.3$. In both of these figures, the left panels are the results for a co-rotating accretion disk, and the right panels are for counter-rotating circulation. The red solid curves and the blue solid lines are the behavior for $\tilde{\xi}\neq 0$ with decreasing values of $\tilde{\xi}$ as read from 
Tables~\ref{tab:1} and \ref{tab:2}, and the black-dashed curves show the behavior for the classical Kerr BH ($\tilde{\xi}=0$). It is clear that in all the cases the quantum gravity effects lead to a lifting of the thermal characteristics of the accretion disk and to a spectrum shifted toward higher frequencies (see the plots of the luminosity in Figs.~\ref{F5} and \ref{F6}).\\

\noindent
Comparing only solid red and black-dashed curves we see that, for a fast rotating BH with $a^{\star}=0.98$ and co-rotating disk, the maximum of the profile of the energy flux from the disk can be as high as $\sim 10^3$ times the flux from a classical Kerr BH, the temperature is approximately $2.3$ times the classical prediction, and the observed luminosity can reach a value that is of the order of $20$ times the GR one. For counter-rotating motion, instead, the increases are very modest with the following approximate factors with respect to the GR results: $1.007$ for the energy flux, $1.001$ for the temperature, and $1.05$ for the luminosity. Note that the increase in luminosity is accompanied for a shift of the spectrum toward higher frequencies.\\

\noindent
For a low spin BH, and comparing again solid red and black-dashed curves, Fig.~\ref{F6} shows that for prograde motion, the peaks of the profiles of the energy flux, the temperature and the observed luminosity, grow approximately by the factors $3.6$, $1.4$ and $\sim 10^3$, respectively. For retrograde motion, the peaks of the profiles of these thermal characteristics increase by the approximate factors $1.4$, $1.1$, and $12$, respectively. Also in this case we have a shift in the frequency spectrum toward higher values.\\

\noindent
It is worth noticing that the quantum effects described are more pronounced for fast and slow rotating RGI-BHs accreting mass from a prograde disk.\\

\noindent
As for the efficiency for conversion of rest mass to radiant energy, Tables~\ref{tab:3} and \ref{tab:4} shows that the RGI-Kerr BH is more efficient than the classical Kerr BH that is, that the efficiency $\epsilon$, calculated from Eq.~(\ref{eq37}), grows as $\tilde{\xi}$ gets larger. The greatest increase in $\epsilon$ occurs for a rapid rotating black hole surrounded by a prograde disk. In this case, Table~\ref{tab:3} shows that for a BH characterized by $a^{\star}=0.98$ and $\tilde{\xi}=0.019$ the accretion efficiency exceeds in about $39\%$ the efficiency for the same spin and $\tilde{\xi}=0$ (the classical case), and at the same time, Table~\ref{tab:1} shows that this increase in $\epsilon$ corresponds to the smaller value $x_{\rm isco}=1.2075$ of the ISCO radius.\\

\begin{table}[!ht]
 \centering
 \begin{ruledtabular}
 \begin{tabular}{p{2.5cm}p{2.5cm}p{2.5cm}|p{2.5cm}p{2.5cm}p{2.5cm}}
 \multicolumn{3}{c}{Co-rotating ($a^{\star}=0.98$, $\tilde{\xi}_{c+}=0.0199$)} & \multicolumn{3}{c}{Counter-rotating ($a^{\star}=0.98$, $\tilde{\xi}_{c+}=0.0199$)}\\ \hline
 $\tilde{\xi}$ & $k_{\rm isco}$ & $\epsilon$ (\%) & $\tilde{\xi}$ & $k_{\rm isco}$ & $\epsilon$ (\%)\\ \hline
 0 & 0.7661 & 23.39 & 0 & 0.96200 & 3.799\\
 0.010 & 0.7410 & 25.90 & 0.010 & 0.96198 & 3.801\\
 0.019 & 0.6742 & 32.48 & 0.019 & 0.96196 & 3.803\\
 \end{tabular}
 \end{ruledtabular}
  \caption{\label{tab:3}The specific energy at the ISCO $k_{\rm isco}$ and the accretion efficiency $\epsilon$ for the same values of $\tilde{\xi}$ in Table~\ref{tab:1} and for $a^{\star}=0.98$. The left column is for prograde motion whereas the right column is for retrograde motion.}
\end{table}
\begin{table}[!ht]
 \centering
 \begin{ruledtabular}
  \begin{tabular}{p{2.5cm}p{2.5cm}p{2.5cm}|p{2.5cm}p{2.5cm}p{2.5cm}}
 \multicolumn{3}{c}{Co-rotating ($a^{\star}=0.3$, $\tilde{\xi}_{c+}=0.5331$)} & \multicolumn{3}{c}{Counter-rotating ($a^{\star}=0.3$, $\tilde{\xi}_{c+}=0.5331$)}\\ \hline
 $\tilde{\xi}$ & $k_{\rm isco}$ & $\epsilon$ (\%) & $\tilde{\xi}$ & $k_{\rm isco}$ & $\epsilon$ (\%)\\ \hline
 0 & 0.9306 & 6.94 & 0 & 0.9508 & 4.91\\
 0.40 & 0.9217 & 7.83 & 0.25 & 0.9494 & 5.06\\
 0.50 & 0.9182 & 8.18 & 0.50 & 0.9477 & 5.22\\
 \end{tabular}
 \end{ruledtabular}
  \caption{\label{tab:4}The specific energy at the ISCO $k_{\rm isco}$ and the accretion efficiency $\epsilon$ for the same values of $\tilde{\xi}$ in Table~\ref{tab:2} and for $a^{\star}=0.3$. The left column is for prograde motion whereas the right column is for retrograde motion.}
\end{table}

\noindent
It is then clear that is the shifting of the ISCO to smaller values that gives rise to the growth of the maximum possible binding energy of the particles in the accretion disk which occurs just at $x_{\rm isco}$ or, equivalently, that gives rise to the increase of the efficiency for conversion of accreted mass into radiation. Obviously, a lower ISCO also results in an increase of the thermal characteristics of the disk.  
%
\section{Conclusions}
In this work, within the AS scenario for quantum gravity, we have addressed the question about the quantum gravity modifications to the thermal properties of a relativistic NPT thin disk accreted by a rotating black hole. The spacetime geometry is determined by the low energy limit of the improved Kerr metric. In this limit, in addition to the usual parameters $M$ and $a^{\star}$, the black hole is also characterized by a free parameter $\tilde{\xi}$ that describes the quantum gravity effects. By varying this parameter in an allowed range dictated by the value of $a^{\star}$, we calculate quantum gravity effects on the time averaged energy flux emitted by the accretion disk, on the disk temperature, on the observed luminosity, and on the efficiency for conversion of mass energy of the accreted material into radiation. Our main goal has been to confront the predictions of the AS theory with the ones of the classical GR both for rapid and slow rotating BH surrounded by prograde and retrograde disks.\\

\noindent
In all the cases analyzed (i.e., fast rotating BH with co-rotating and counter-rotating disk, and slow rotating BH with co-rotating and counter-rotating disk), we have found that growing values of $\tilde{\xi}$ displace the ISCO toward smaller values which leads to an increase of the maximum of the profiles of all the radiation properties of the disk, to a shift of the spectrum toward higher frequencies, and to a greater accretion efficiency. These quantum effects are more conspicuous for a co-rotating disk spiraling around high and low spin black holes. Except for the natural distinction among prograde and retrograde disk, these results are in agreement with the previously announced for an accretion disk around a RGI-Schwarzschild BH in Ref.~\cite{r17}.\\

\noindent
We must notice that the quantum gravity effects on the thermal properties of an accretion disk around RGI-Schwarzschild and Kerr black hole that we find in \cite{r17} and in the present work, are not exclusive of the AS theory. Indeed, similar predictions arise, for example, in: (a) $f(R)$ gravity which is a classical extension of GR that adds a function $f(R)$ of the Ricci scalar $R$ to the Einstein-Hilbert action. In Ref.~\cite{r30}, the authors show that in the strong field regime and for constant Ricci scalar $R=R_0$ in the allowed range $-1.2 \times 10^{-3}\leq R_0 \leq 6.67 x 10^{-4}$, the radius of the ISCO decreases and, correspondingly, the peaks of the temperature and luminosity of an NPT disk increase provided $R_0<0$ as compared with the peaks in the case of the classical Kerr solution. For $R_0>0$, the radiation properties of the disk have no significant differences with the classical case; (b) In 4D-Einstein-Gauss-Bonnet (4EGB) gravity, which is also a classical extension of GR, where the coupling $\alpha$ to the Gauss-Bonnet term in the action is re-scaled as $\alpha \rightarrow \alpha/(D-4)$ in such a way that the Gauss-Bonnet invariant can make non-trivial contribution to the gravitational dynamics in the limit $D\rightarrow 4$ \cite{r31}. In the context of this extension, in \cite{r32}, it has been shown that for positive values of $\alpha$, the relativistic thin accretion disk around the static spherically symmetric black hole is hotter, more luminosity, and more efficient than the one around a classical Schwarzschild black hole with the same mass, while it is cooler, less luminosity, and less efficient for negative values of this coupling. These same results are found when a thin disks around a rotating 4EGB black hole is considered \cite{r33}; (c) In models of regular BH as the static spherically symmetric Bardeen BH and the Hayward BH studied in \cite{r34} where it is concluded that, for the values of the parameters $l_b$ and $l_h$ appearing in the mass function of the Bardeen and Hayward models, respectively, such that $l_b=l_h \leq l_c$ (where $l_c=4/(3\sqrt{3})$ is the critical value below which a black hole solution exists), the energy radiated from the surface of the disk, its temperature, luminosity and efficiency increase as a result of the shifting of the ISCO radius toward smaller values. The increases are greater for the Bardeen BH than for the Hayward BH; (d) In the analysis of the orbits of a spinning test particle moving around the RGI-Schwarzschild and Kerr BH proposed in \cite{r1}, which shows that the ISCO decreases under an increase of the spin $s$ of the test particle \cite{r35}, and (e) In GR coupled to nonlinear electrodynamics (NLED), where the study of the thermal properties of a thin disk surrounding a BH with magnetic charge $q_m>0$ and mass function depending of the NLED coupling $\beta>0$ performed in \cite{r36,r37}, indicates that an increase in $\beta$ results in an increase in the energy flux, temperature and luminosity of the disk, while it shows a completely opposite behavior as the magnetic charge $q_m$ increases. The change in the luminosity spectrum for increasing $q_m$ is more prominent than for increasing $\beta$ with respect to the classical Schwarzchild black hole.\\

\noindent
From these examples, we can conclude that the findings we are reporting in this work are not sufficient to distinguish a rotating RGI-black hole from rotating black holes in other theoretical extensions of GR proposed in the literature. However, alongside with the studies based on the iron line shape, X-ray reflection spectroscopy, and shape and size of the BH shadow, our results contribute to accumulate information that can be useful to probe the AS gravity theory with future observational evidences.

\section*{ACKNOWLEDGEMENTS}
We acknowledge financial support from Universidad Nacional de Colombia.

\end{document}